\documentclass[12pt]{article}

\usepackage{times}
\usepackage{graphics}
\usepackage{amssymb}
\usepackage{amsmath}

\begin{document}

\baselineskip 6mm
\renewcommand{\thefootnote}{\fnsymbol{footnote}}


\newcommand{\nc}{\newcommand}
\newcommand{\rnc}{\renewcommand}


\rnc{\baselinestretch}{1.24}    
\setlength{\jot}{6pt}       
\rnc{\arraystretch}{1.24}       

\makeatletter
\rnc{\theequation}{\arabic{equation}}
\@addtoreset{equation}{section}
\makeatother



\nc{\be}{\begin{equation}}
\nc{\ee}{\end{equation}}
\nc{\bea}{\begin{eqnarray}}
\nc{\eea}{\end{eqnarray}}
\nc{\xx}{\nonumber\\}

\nc{\eq}[1]{(\ref{#1})}
\nc{\newcaption}[1]{\centerline{\parbox{6in}{\caption{#1}}}}

\nc{\fig}[3]{
\begin{figure}
\centerline{\epsfxsize=#1\epsfbox{#2.eps}}
\newcaption{#3. \label{#2}}
\end{figure}
}


\nc{\np}[3]{Nucl. Phys. {\bf B#1} (#2) #3}
\nc{\pl}[3]{Phys. Lett. {\bf #1B} (#2) #3}
\nc{\prl}[3]{Phys. Rev. Lett.{\bf #1} (#2) #3}
\nc{\prd}[3]{Phys. Rev. {\bf D#1} (#2) #3}
\nc{\ap}[3]{Ann. Phys. {\bf #1} (#2) #3}
\nc{\prep}[3]{Phys. Rep. {\bf #1} (#2) #3}
\nc{\rmp}[3]{Rev. Mod. Phys. {\bf #1} (#2) #3}
\nc{\cmp}[3]{Comm. Math. Phys. {\bf #1} (#2) #3}
\nc{\mpl}[3]{Mod. Phys. Lett. {\bf #1} (#2) #3}
\nc{\cqg}[3]{Class. Quant. Grav. {\bf #1} (#2) #3}
\nc{\jhep}[3]{J. High Energy Phys. {\bf #1} (#2) #3}


\def\vare{\varepsilon}
\def\bz{\bar{z}}
\def\bw{\bar{w}}


\def\CA{{\cal A}}
\def\CC{{\cal C}}
\def\CD{{\cal D}}
\def\CE{{\cal E}}
\def\CF{{\cal F}}
\def\CG{{\cal G}}
\def\CI{{\cal I}}
\def\CT{{\cal T}}
\def\CM{{\cal M}}
\def\CN{{\cal N}}
\def\CP{{\cal P}}
\def\CL{{\cal L}}
\def\CV{{\cal V}}
\def\CS{{\cal S}}
\def\CW{{\cal W}}
\def\CY{{\cal Y}}
\def\CS{{\cal S}}
\def\CO{{\cal O}}
\def\CP{{\cal P}}
\def\CN{{\cal N}}


\def\IC{\mathbb{C}}
\def\ID{\mathbb{D}}
\def\IH{\mathbb{H}}
\def\IP{\mathbb{P}}
\def\IR{\mathbb{R}}
\def\IZ{\mathbb{Z}}


\def\a{\alpha}
\def\b{\beta}
\def\ga{\gamma}
\def\d{\delta}
\def\ep{\epsilon}
\def\ph{\phi}
\def\k{\kappa}
\def\l{\lambda}
\def\m{\mu}
\def\n{\nu}
\def\th{\theta}
\def\rh{\rho}
\def\s{\sigma}
\def\t{\tau}
\def\w{\omega}
\def\G{\Gamma}


\def\half{\frac{1}{2}}
\def\imp{\Longrightarrow}
\def\dint#1#2{\int\limits_{#1}^{#2}}
\def\goto{\rightarrow}
\def\para{\parallel}
\def\brac#1{\langle #1 \rangle}
\def\del{\nabla}
\def\grad{\nabla}
\def\curl{\nabla\times}
\def\div{\nabla\cdot}
\def\p{\partial}
\def\e{\epsilon_0}


\def\Tr{{\rm Tr}}
\def\det{{\rm det}}
\def\im{{\rm Im~}}
\def\re{{\rm Re~}}


\def\Kahler{K\"{a}hler}


\def\e{\varepsilon}
\def\bA{\bar{A}}
\def\c{\zeta}

\begin{titlepage}
\hfill\parbox{4cm}
{hep-th/0011233 \\KIAS-P00076}

\vspace{15mm}
\centerline{\Large \bf Factorization and generalized $*$-products}

\vspace{10mm}
\begin{center}
Youngjai Kiem, Dong Hyun Park\footnote{ykiem,
donghyun@newton.skku.ac.kr}
\\[2mm]
{\sl BK21 Physics Research Division and Institute of Basic
Science}
\\
{\sl Sungkyunkwan University, Suwon 440-746, Korea}
\\[5mm]
Sangmin Lee\footnote{sangmin@kias.re.kr}
\\[2mm]
{\sl School of Physics, Korea Institute for Advanced Study, Seoul
130-012, Korea}
\end{center}

\thispagestyle{empty}
\vskip 40mm

\centerline{\bf ABSTRACT}
\vskip 5mm
\noindent
The generalized $*$-products, or the $*_N$-products, appear both
in the one-loop effective action of noncommutative Yang-Mills
theories and in the coupling of a closed string to $N$ open
strings on a disk when the D-brane world-volume is noncommutative.
Factorization of the string amplitudes provides a uniform
understanding of the $*_N$-products and a hint to obtain a simple,
explicit formula (in the momentum space) for arbitrary $N$ in the
non-Abelian case. Possible extension to a more general ${}_M
*_{N}$-product in the $M$-loop context is discussed.

\vspace{2cm}
\end{titlepage}

\baselineskip 7mm
\renewcommand{\thefootnote}{\arabic{footnote}}
\setcounter{footnote}{0}

Field theories on noncommutative $\IR^n$ (NCFT) \cite{connes,filk}
can be obtained by replacing the ordinary product between two
fields with the $*$-product defined as
\begin{equation}
 \phi_1 * \phi_2 (x) = \exp \left[ \frac{i}{2} \theta^{\mu \nu}
\frac{\partial}{\partial y^{\mu}} \frac{\partial}
 {\partial z^{\nu}} \right] \phi_1 (y ) \phi_2 (z) |_{y=z=x} ~ .
\label{star}
\end{equation}
By taking an appropriate decoupling limit where the string length
scale goes to zero, the noncommutative field theories are obtained
from string theory with a constant background NS-NS two-form field
\cite{string,witten}. In particular, the $*$-product in
(\ref{star}) is reproduced from the tree-level vertex operator
product \cite{witten}
\begin{equation}
\label{ope1}
 \exp \left(  i p_1 X (x)  \right) \exp \left(  i p_2 X
 (x^{\prime} ) \right) \simeq \exp \left( - p_{1 \mu} \langle X^{\mu} (x)
 X^{\nu} ( x^{\prime} ) \rangle p_{2 \nu} \right) ~ ,
\end{equation}
where we suppressed the polarization dependent part of the vertex
operators, with
\begin{equation}
\langle X^{\mu} (x)  X^{\nu} ( x^{\prime} ) \rangle \simeq
  \frac{i}{2} \theta^{\mu \nu} \epsilon ( x - x^{\prime} )
 + \cdots .  \label{ope}
\end{equation}
The function $\epsilon ( x - x^{\prime} )$ is the Heaviside step
function.  In \eqref{ope1}, the part of the world-sheet propagator
that depends only on the ``ordering'' of the vertex operators is
retained.  Since every vertex operator in any string theory
contains the piece shown in (\ref{ope}), the appearance of the
$*$-product structure is quite generic and universal.

At the loop level\footnote{
Bearing in mind that a string diagram reduces to a set of field
theory diagrams in the decoupling limit, we will use the terms
`tree,' `one-loop' and so on interchangeably with `disk,'
`annulus,' etc.}, NCFT exhibit much richer physics, such as the
UV/IR mixing
\cite{uvir,oneloop,twoloop}. One of the recent additions to the
list of intriguing loop physics is the appearance of the
generalized $*$-products in the non-planar one-loop amplitudes
\cite{liu}.
In the context of the NCFTs, these generalized $*$-products appear
in the one-loop double trace terms in the effective action
\cite{liu,zanon}, in certain $U(1)$ anomaly computations
\cite{ardalan} and in the expansion of open Wilson line 
\cite{openw,mehen}.
They also play an essential role in the
construction of the solutions of the Seiberg-Witten map and in the
issue of the gauge invariance in nonlocal theories
\cite{mehen,liu2,garous,das,garousi}. 
See also \cite{misra}.
Furthermore, in an
apparently unrelated context of the closed string absorption
amplitudes at the disk level, the same kind of generalized
$*$-products appear \cite{hyun,garous,das,garousi}.

The main theme of the present note is to uniformly understand the
appearance of the generalized $*$-products in various contexts by
studying the factorization limits of the string amplitudes. This
line of approach was first suggested in Ref.~\cite{hyun} for a
closed string coupling to two open string
states\footnote{In
\cite{verlinde}, it was observed that the bulk closed string
vertex insertions and the open string loops can be treated on an
equal footing in the factorization limit.}. For example, the fact
that both the closed string absorption amplitudes and the one-loop
double trace terms in the effective action of Yang-Mills theory
contain the same generalized $*$-product structures is natural
from our analysis\footnote{Recently, from the analysis of the
noncommutative Born-Infeld action, an alternative explanation for
the same fact was reported in \cite{das}.}. In this process, we
obtain a simple and explicit expression for the $*_N$-products in
a closed form for arbitrary $N$ in the non-Abelian context.
Furthermore, our analysis indicates that, in the case of the
multi-loop amplitudes in the open string context and in the case of
the multiple closed string absorption amplitudes on a disk, there
might be yet another generalized class of $*$-products.

Our presentation does not sensitively depend on the existence of
supersymmetry; when there are supersymmetries, some terms that we
write down vanish as the cancellation between the space-time
bosonic and fermionic contributions occurs. Each contribution,
however, contains the generalized $*$-product parts. In addition,
our analysis should be applicable to space-time, space-space and
light-like noncommutative field/string theories \cite{lightlike}.

\begin{figure}[ht]
\label{fig1}
\centerline{\scalebox{0.9}{\includegraphics{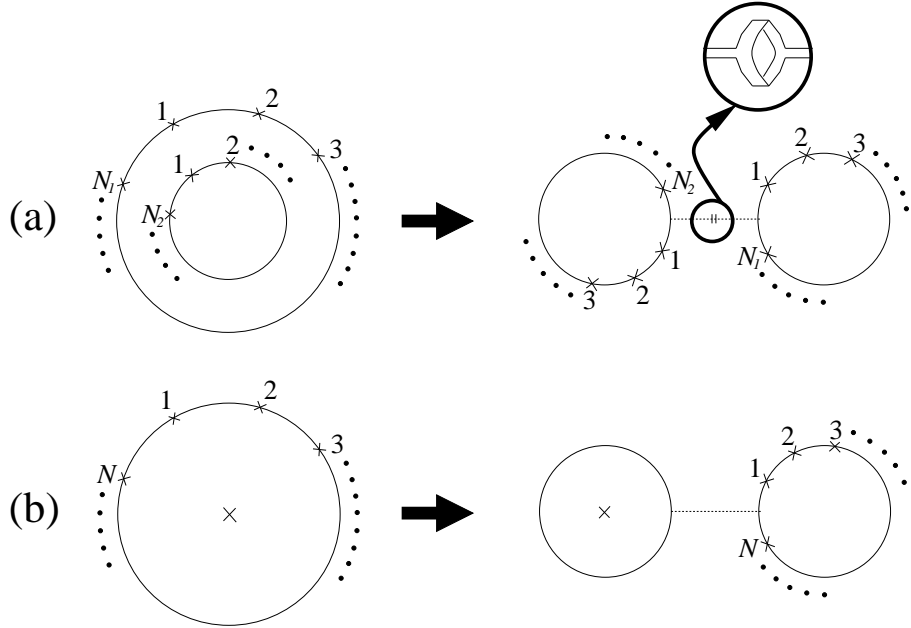}}}
\caption{The factorization limits for the open string
annulus amplitudes and the disk closed string absorption
amplitude.  The double twisted line should be used
for the connecting leg in the case of annulus amplitude.}
\end{figure}

The consideration of the factorization limit is helpful in
understanding the reason why the coupling of a single closed
string with $N$ open strings gives the same $*_N$-product
structure as the one-loop amplitude involving purely open string
insertions.  The $*$-product that appears at the tree level and in
the planar multi-loop amplitudes
\cite{klp} does not depend on the moduli of the world-sheet and
the local position of the open string vertex insertions, but only
depends on the ordering of the vertex insertions along a connected
boundary. Keeping this in mind, we first consider the open string
one-loop amplitude shown in Fig.~1(a).

The case of interest is the factorization limit of the amplitude
where the boundary inside approaches the outer boundary.  The
amplitude in this limit factorizes into the product of two disk
amplitudes, each of which hosting the external open string vertex
insertions. The connecting leg in this case represents the
momentum flowing between two different boundaries through a
`mini-annulus.' As we will find out soon, the object that should be
computed to produce the $*_N$-products is the moduli independent
part of the disk amplitude with a connecting leg inserted as shown
in Fig.~2.

Next, we consider the single closed string absorption amplitude,
corresponding to Fig.~1(b).  We suppose that enough number of open
string vertices are inserted to fix the $SL(2,\IR)$ invariance of
the disk world-sheet. The relevant factorization limit in this
case is the limit where the bulk closed string vertex approaches
the disk boundary. The amplitude in consideration becomes the
product of a disk amplitude with a bulk closed string insertion
and another disk amplitude with $N$ boundary insertions. In this
factorization limit, no `mini-annulus' is involved. However, the
closed string insertion can be regarded as the specification of
the boundary state along the `blown-up' insertion point. As the
incoming closed string momentum passes through the `cylinder'
(which corresponds to the disk on the left hand side of
Fig.~1(b)), it becomes topologically similar to the
`mini-annulus.'  Noting that the $*_N$-product comes from the
moduli-independent part of the amplitudes, which does not change
under the change of the moduli, we again see that the relevant
object is the disk amplitude with a connecting leg insertion in
Fig.~2.

\begin{figure}[ht]
\label{fig2}
\centerline{\scalebox{1}{\includegraphics{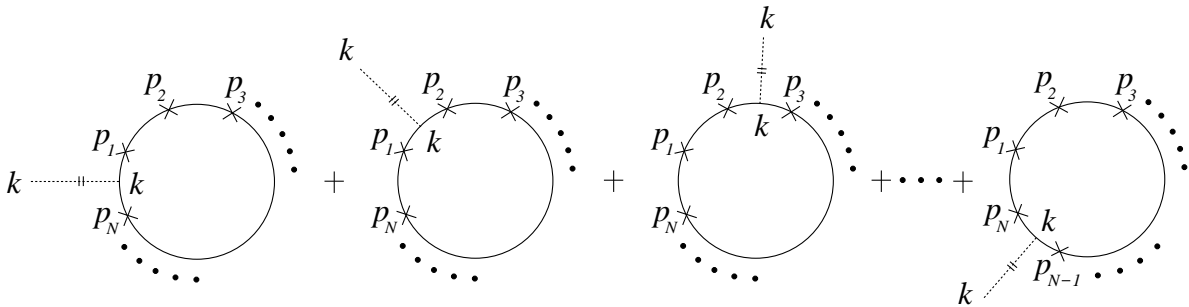}}}
\caption{ All $N$ cyclic permutations, which have
the same Chan-Paton factor, should be added.
The first term in the summation in the figure is computed
in (\ref{phase}).  From the factorization point of view,
the summation is automatically generated for a given
fixed ordering of the vertices along the disk boundary.}
\end{figure}

To obtain the explicit expression for the $*_N$-products, we
compute the object in Fig.~2 from the open string point of view.
As it should, the computation gives the identical answer to the
one obtained by the computation of the single closed string
absorption amplitude \cite{garousi}.  In Fig.~2, there are $N$
open string vertex insertions along the boundary of the disk.
There are $N! = N \times (N-1)!$ permutations of the external
vertices; among these $N!$ permutations, the $N$ cyclic
permutations do not change the Chan-Paton factor ${\rm Tr} (T_1
\cdots T_N)$. These cyclic permutations are summed in Fig.~2.
In fact, from the factorization point of view, the summation over
the cyclic permutations is automatically generated for a fixed
ordering of the vertices along the disk boundary; the connecting
leg can be attached to any segment between the two external
vertices insertions (corresponding to different factorization
channels). On the other hand, non-cyclic permutations change the
Chan-Paton factor (assuming that the number of D-branes are large
enough) and should be considered independently.

{}From \cite{klp}, the one-loop planar and non-planar open string
propagators are given by
\begin{eqnarray}
\label{pprop1}
G^{\m\n}_P (z,z') &=&  \a' G^{\mu \nu} G(z, z') + \frac{i}{2}
\theta^{\m\n} \e(z-z'),
\\
G^{\m\n}_{NP} (z,z') &=& \a' G^{\mu \nu} G(z, z') - \displaystyle
\frac{(\theta G \theta )^{\mu \nu}}{2\pi\a' T}  (x-x')^2
- \frac{2 i}{T} \theta^{\m\n} (x-x')(y + y'),
\label{npprop1}
\end{eqnarray}
where the function $G(z,z')$ is defined as
\begin{equation}
G(z,z') =  -\log \left|
\frac{\theta_1(z-z'|iT)}{\theta'_1(0|iT)} \right|^2 +
\frac{2\pi}{T} (y-y')^2 ~.
\end{equation}
Here the function $\e(z-z' ) = {\rm sgn} ( y - y^{\prime})$ for
the boundary at $x = 0$ and
$\e(z-z' ) = - {\rm sgn} ( y - y^{\prime})$ for the boundary
at $x = 1/2$.
The momentum $k$ flowing along the connecting leg, which
is identical to the momentum flowing between two boundaries,
satisfies the momentum conservation condition
\begin{equation}
 k + p_1 + \cdots + p_N = 0 ~ .
\label{conserv}
\end{equation}
At the level of the scattering amplitudes ($\exp ( - p_{\mu}
G^{\mu \nu} p_{\nu} ) $), there are non-planar contribution coming
from the contractions between the free edge point (located at $x =
0$ and $y = 0$) and the points along the disk (located at
$x^{\prime} = 1/2$ and $y^{\prime} = \tau_i T)$. Furthermore,
among the points along the disk, there are planar contributions.

Collecting the terms of the order of $T^0$, which are
moduli-independent, from (\ref{pprop1}) and (\ref{npprop1}), we
have (for the particular ordering of the external vertices as
shown in the first term of Fig.~2):
\begin{equation}
\CA_{P} \CA_{NP}=
 \exp \left[  \frac{i}{2} \sum_{i< j} p_i \times p_j  \right]
\int_0^1 d \tau_1 \int_0^{\tau_1} d \tau_2 \cdots
\int_0^{\tau_{N-1}} d \tau_N  \exp \left[ - i \sum_{i=1}^{N}
 k \times p_i ~ \tau_i  \right] ~ ,
\label{phase}
\end{equation}
where the first exponential term $\CA_P$ originates from the
planar contribution, and the second exponential term $\CA_{NP}$
comes from the non-planar contribution\footnote{H.~Liu informed
us that the particular form of $*_N$-products given in
(\ref{phase}) is the one that naturally comes from
the open Wilson lines.  See Eqs.~(2.6)-(2.13) of \cite{liu2}.}.
When writing down
(\ref{phase}), one should keep in mind that
\begin{equation}
  \left( \sum_{i =1}^N p_i \right) \times k = - k \times k = 0
\end{equation}
in view of the momentum conservation condition (\ref{conserv}).
Noting the identities
\begin{equation}
  \sum_{i < j} p_i \times p_j ( \tau_i - \tau_j )
 =  \frac{1}{2} \sum_{i, j}  p_i \times p_j ( \tau_i - \tau_j )
 = - \sum_{i,j} p_j \times p_i ~ \tau_i
 =  \sum_i k \times p_i ~ \tau_i
\end{equation}
we find that (\ref{phase}) is identical to the definition given in
\cite{liu}.  Written in the form of (\ref{phase}), it is obvious
that when there is no momentum flow between the two boundaries, $k
= 0$, the conventional planar phase term defining the ordinary
$*$-product is enough; in particular, this situation includes the
case of the planar vertex insertions or, equivalently, the single
trace terms in the effective action.

The integration involved in $\CA_{NP}$
can be evaluated to yield:
\begin{subequations}
\label{integral}
\begin{eqnarray}
 \CA_{NP} ( p_1 , \cdots , p_N ) &=&
  \CA_{NP}^1 ( p_1 , \cdots , p_N )
 + \CA_{NP}^2 ( p_1 , \cdots , p_N )
\\
\label{hana}
\CA_{NP}^1 &=& \dfrac{1}{\prod_{i=2}^{N}
\left( -i k \times P_i \right) }
\\
\label{dool}
\CA_{NP}^2 &=&
\sum_{\epsilon}{}' ( - 1 )^{\sum_{i=2}^{N} \epsilon_i }
 \dfrac{ 1 - \exp \left( - i k \times q_1^{\epsilon} \right) }
 { \prod_{i=1}^N \left( i k \times
  q^{\epsilon}_i \right) } ~ ,
\end{eqnarray}
\end{subequations}
The various objects appearing in (\ref{integral}) are as follows.
For efficient bookkeeping, we introduced an $(N-1)$ objects
$\epsilon = (\epsilon_2 , \cdots , \epsilon_N )$ with each
$\epsilon_i = 0$ or $1$. The primed summation in
\eqref{dool} is over all possible values of $\epsilon$ except the
case $\epsilon_2 =\cdots = \epsilon_N = 1$, which we separated in
\eqref{hana}. The $N$ objects $q^{\epsilon} = (q^{\epsilon}_1 ,
\cdots , q^{\epsilon}_N )$ is the following linear combination of
$p_i$'s;
\begin{equation}
q^{\epsilon}_i =
\sum_{j=i}^N \left( \prod_{k=i+1}^{j} \epsilon_{k} \right) p_{j}
= p_i + \epsilon_{i+1} p_{i+1}
 + \epsilon_{i+1} \epsilon_{i+2} p_{i+2} + \cdots
 + \epsilon_{i+1} \epsilon_{i+2} \cdots \epsilon_N p_N ~ ,
\end{equation}
In particular, this implies that
\begin{equation}
 q^{\epsilon}_1 = p_1 + \epsilon_{2} p_2
 + \epsilon_{2} \epsilon_{3} p_3 + \cdots
 + \epsilon_{2} \epsilon_{3} \cdots \epsilon_N p_N
\quad
{\rm and}
\quad
P_i \equiv q^{(1, 1, \cdots , 1)}_i = \sum_{j=i}^N p_{i} ~ .
\end{equation}
Remarkably, the complicated $\CA^2_{NP}$ term in \eqref{dool}
completely cancels out when summed over $N$ cyclic permutations.
That is,
\begin{equation}
 \CA_{P}( p_1 , \cdots , p_N ) \CA_{NP}^2 ( p_1 , \cdots , p_N ) +
 ( {\rm cyclic} ~ {\rm permutations} )  =  0 ~.
 \label{thm}
\end{equation}
For the proof, it is convenient to classify the terms in the
summation by the value of $|\epsilon| \equiv \sum_{i=1}^{N-1}
\epsilon_i$. For each value of $|\epsilon|$ ranging from $0$ to $N-2$
($|\epsilon| = N-1$ is excluded from the summation), there are
$_{N-1} C_{|\epsilon|}$ terms; all the terms having the same
$|\epsilon|$ have the same sign $(-1)^{|\epsilon|}$ in
$\CA_{NP}^2$. As for the $e^{- i k \times q^{\epsilon}_1}$ term in
the numerator, note that $q^{\epsilon}_1 = p_1 + \cdots + p_k$ if
$k$ is the smallest integer satisfying
$\epsilon_2 = \cdots = \epsilon_k = 1$ and $\epsilon_{k+1} = 0$.
When multiplied by this exponential factor, the planar phase gets
shifted by $k$ steps in cyclic permutation:
\begin{equation}
\begin{array}{rcl}
\CA_P(p_1, \cdots p_N) \cdot e^{-ik\times (p_1 + \cdots + p_k)}
&=&
\CA_P (-p_1, \cdots, -p_k, p_{k+1}, \cdots p_N)
\\
&=& \CA_P (p_{k+1},\cdots, p_N, p_1, \cdots, p_k).
\label{ggao}
\end{array}
\end{equation}
Going to the second line of (\ref{ggao}), we use the property
that the $k$ step cyclic permutation for the
$\CA_P$ corresponds to the transformation
$p_1 , \cdots , p_k \rightarrow - p_1 , \cdots , - p_k$.
By writing out the denominators and comparing them among different
cyclic permutations, one can show that the $e^{- i k \times
q^{\epsilon}_1} \CA_{P} $ terms cancel the $-1 \times \CA_{P}$
terms completely for each value of $|\epsilon|$. Take $N=3$ and
$|\epsilon|=2$ for example. Up to an overall normalization, the
$e^{- i k \times q^{\epsilon}_1} \CA_{P}$ terms are
\begin{eqnarray}
\label{cancel1}
\begin{array}{rcl}
&(123):& \quad
\frac{\CA_F (312)}{k\times(p_1+p_2) ~ k\times p_2 ~ k\times p_3} +
\frac{\CA_F (231)}{k\times p_1 ~ k\times (p_2+p_3) ~ k\times p_3}
\\
&(231):& \quad
\frac{\CA_F (123)}{k\times(p_2+p_3) ~ k\times p_3 ~ k\times p_1} +
\frac{\CA_F (312)}{k\times p_2 ~ k\times(p_3+p_1) ~ k\times p_1}
\\
&(312):& \quad
\frac{\CA_F (231)}{k\times(p_3+p_1) ~ k\times p_1 ~ k\times p_2} +
\frac{\CA_F (123)}{k\times p_3 ~ k\times (p_1+p_2) ~ k\times p_2},
\end{array}
\end{eqnarray}
while the $(-1)$ terms are
\begin{eqnarray}
\label{cancel2}
\begin{array}{rcl}
&(123):& \quad -
\frac{\CA_F (123)}{k\times(p_1+p_2) ~ k\times p_2 ~ k\times p_3} -
\frac{\CA_F (123)}{k\times p_1 ~ k\times (p_2+p_3) ~ k\times p_3}
\\
&(231):& \quad -
\frac{\CA_F (231)}{k\times(p_2+p_3) ~ k\times p_3 ~ k\times p_1} -
\frac{\CA_F (231)}{k\times p_2 ~ k\times (p_3+p_1) ~ k\times p_1}
\\
&(312):& \quad -
\frac{\CA_F (312)}{k\times(p_3+p_1) ~ k\times p_1 ~ k\times p_2} -
\frac{\CA_F (312)}{k\times p_3 ~ k\times (p_1+p_2) ~ k\times p_2}.
\end{array}
\end{eqnarray}
Clearly, every single term in \eqref{cancel1} cancels with a term
in \eqref{cancel2}.

Using (\ref{thm}), one can
thus write the following simple expression for the
$*_N$-products in the momentum space:
\begin{eqnarray}
{\rm Tr } *_N \left[ f_1 (p_1 ) , f_2 (p_2) , \cdots
 f_N ( p_N ) \right] =  \sum_{(N-1)! } f^{a_1}_1 (p_1 )
 \cdots f_N^{a_N} (p_N )
 {\rm Tr} \left( T^{a_1} \cdots T^{a_N} \right) \nonumber \\
\times \left( \frac{ \exp \left[  \frac{i}{2}
 \sum_{i< j} p_i \times p_j \right]}
 { \prod_{i=2}^{N} \left( -i k \times P_i \right) }
 + ({\rm cyclic} ~ {\rm permutations } ) \right)
\label{theresult}
\end{eqnarray}
where the summation runs over the independent Chan-Paton factor,
$f_i = \sum_{a_i} f_i^{a_i} T^{a_i}$, $T^{a_i}$ are generators of
the Chan-Paton group $U( n )$ for $n$ D-branes, and $k = - ( p_1 +
\cdots + p_N )$.  Written explicitly for $N=3$, for example, ${\rm
Tr } *_3 \left[ , ,  \right]$ contains
\begin{equation}
 {\rm Tr} \left( T^{a_1} T^{a_2} T^{a_3}  \right)
  \left( \frac{ e^{ \frac{i}{2} ( p_1 \times p_2 + p_1 \times p_3
 + p_2 \times p_3 ) } }
 { - ( k \times p_3  ) ( k \times ( p_2 + p_3 ) ) }  +
 ({\rm cyclic}) \right)
 + ( 1 \leftrightarrow 2 ) ~ ,
\end{equation}
where $- k = p_1 + p_2 + p_3 $, which is identical to Eq.~(3.17) of
\cite{liu}. For $N=4$, we get
\begin{eqnarray}
&{\rm Tr} \left( T^{a_1} T^{a_2} T^{a_3} T^{a_4} \right)&
  \left( \frac{ e^{ \frac{i}{2} ( p_1 \times p_2 + p_1 \times p_3
 + p_1 \times p_4 + p_2 \times p_3 + p_2 \times p_4 +
 p_3 \times p_4) } }
 { i ( k \times p_4  ) ( k \times ( p_3 + p_4 ) )
      ( k \times ( p_2 + p_3 + p_4 ) ) }  +
 {\rm (cyclic)} \right)
\\
&&+ ( 5 ~ {\rm noncyclic} ~ {\rm permutations}  ).
\nonumber
\end{eqnarray}
For general value of $N$, there are $N!$ terms in the definition
of $*_N$-products.  In
Ref.~\cite{liu2}, a descent relation between $*_N$ and $*_{N+1}$ is
given in Eq.~(2.18).  It is straightforward to show that
\eqref{theresult} satisfies the descent relation. Reversing the
logic, it might be possible to derive \eqref{theresult} from the
descent relation.

Returning to the open string amplitude shown
in Fig.~1(a), we find that the effective action from that amplitude
contains the $*_N$ products (by computing the two diagrams of the
type shown in Fig.~2 and multiplying them)
\begin{equation}
 {\rm Tr } *_{N_1} \Big[ \underbrace{F , \cdots  , F}_{N_1} \Big]
 {\rm Tr } *_{N_2}
  \Big[ \underbrace{F , \cdots ,  F}_{N_2} \Big] ~ ,
\label{ff1}
\end{equation}
when written schematically.  By the similar token, the
closed string absorption amplitude in Fig.~1(b) yields a term
\begin{equation}
 h {\rm Tr } *_N
 \Big[ \underbrace{F , \cdots ,  F}_N \Big]
\label{ff2}
\end{equation}
among the terms in the effective action,
where $h$ is the closed string field.
A cautionary remark should be in order\footnote{We are grateful
to H.~Liu and H.-T.~Sato for raising this issue.}.  In general,
the intermediate modes running through the connecting leg,
involved in the factorization, contain
infinite number of modes.  The schematic factorized forms
in (\ref{ff1}) and (\ref{ff2}) appear for each intermediate mode.
When summed over all intermediate modes, however, one generically expect
that more complicated (non-factorized) structures arise (see, for example,
Eqs.~(3.2) and (3.26) of Ref.~\cite{liu}).  In this sense, the terms
shown in (\ref{ff1}) and (\ref{ff2}) are small subsets of
terms that can appear in the one-loop effective action,
and in the contact terms in the single closed string
absorptions on a disk, respectively.

\begin{figure}[ht]
\label{fig3}
\centerline{\scalebox{0.8}{\includegraphics{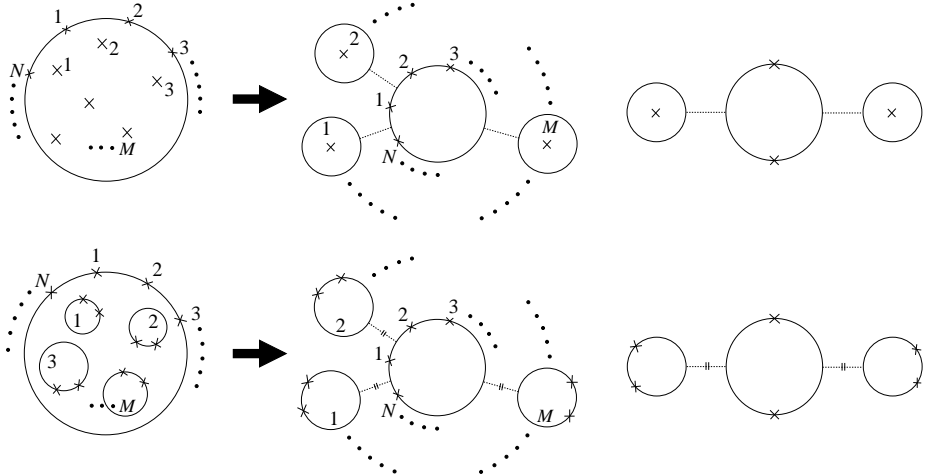}}}
\caption{The schematic depiction of the factorizations involving
${}_M *_{N}$-products, which can arise in the disk amplitudes
involving $M$ bulk closed string insertions or $M$-loop terms in
the open string effective action.}
\end{figure}

Up to now, our consideration has been restricted to the purely
open string insertions at the one-loop level and the tree-level
single closed string absorptions. In the $M$-loop ($M>1$) context
for purely open string insertions, in the case of multiple number
of closed string absorptions at the tree-level, and in the mixed
cases, however, we have more intriguing possibilities. These
situations and their interesting factorization limits are depicted
in Fig.~3. As illustrated in the middle column of Fig.~3, as many
as $M$ connecting legs can be attached to a disk. Generalizing the
combinatorics of Fig.~2 to the case of $M$ connecting legs, one
now should sum over $(M+N-1)!$ diagrams, suggesting the
possibility  of ${}_M *_N$ products.  Some or all of these
products might actually be expressible in terms of $*_N$ products;
still, it remains an open question to see if the $*_N$ products
can be genuinely further generalized.  The simplest nontrivial
examples of the ${}_M *_N$ products arise from the diagrams in the
third column of Fig.~3; the case of two closed string absorptions
on a disk and the two-loop triple trace $ ({\rm Tr} F^2 )^3$ terms
in the effective action of the supersymmetric Yang-Mills theory.
Evaluation of such diagrams and more detailed understanding of
${}_M *_N$ products will be reported elsewhere
\cite{future}.

\section*{Acknowledgements}

Y.~K. would like to thank Jin-Ho Cho, Yoonbai Kim, Phillial Oh and
Haru-Tada Sato for useful discussions.  Y.~K. is also
grateful to Sumit Das for encouragements and to Hong Liu for
interesting comments.


\end{document}